Enlargement of Grains of Silica Colloidal Crystals by Centrifugation in an Inverted-Triangle Internal-Shaped Container

Kaori Hashimoto[1], Atsushi Mori[2], Katsuhiro Tamura[2], and Yoshihisa Suzuki[*,2]

[1]Graduate School of Advanced technology and Science, The University of Tokushima, Tokushima 770-8506, Japan

[2]Institute of Technology and Science, The University of Tokushima, Tokushima 770-8506, Japan

**Abstract**

We successfully fabricated large grains of silica colloidal crystals in an inverted-triangle internal-shaped container (inverted-triangle container) by centrifugation. The largest grain in the container was much larger than that in a container which has a flat bottom and constant width (flat-bottomed container). The edged bottom of the inverted-triangle container eliminated the number of the grains, and then the broadened shape of the container effectively widened the grains.

E-mail : suzuki@chem.tokushima-u.ac.jp

A close-packed fcc silica colloidal crystal was used as a template for the fabrication of an inverse opal with a three dimensional (3D) full photonic band gap (PBG)[1], and it was easily fabricated by drying almost close-packed (with high volume fraction of particles) large columnar-shaped grains of colloidal crystals which are prepared by centrifugation.[2] An inverse opal with 3D full PBG is one of the most promising materials which enable us to fabricate very large-scale 3D photonic integrated circuits (PICs) in the interior of the opal or optical computer in the future.

Although such a large-scale 3D PIC has not yet been realized at this stage, small-scale PICs on several semiconductors have already been produced by laser microfabrication processes[3,4], and laser microfabrication techniques of optical waveguides in close-packed colloidal crystals have also been developed.[5-7] Actually, arbitrary linear dislocations as optical wave guides were induced in close-packed colloidal crystals simply by irradiation of laser beams.[5,7] In particular, Taton et al., showed that three-dimensional optical waveguides were successfully produced via irradiation of laser beams into the colloidal crystals.[7] This method will be useful for the fabrication of a very large-scale 3D PIC into a bulky grain of close-packed colloidal crystals. They also suggested that the process with colloidal crystal will be appropriate for mass fabrication of PICs, since the crystal is easily obtained via self-assembly processes. In addition to their suggestion, we'd like to stress the potential of a large grain for the high yield of small PIC devices which are produced by cutting out of the grain.

Of course, in the case of non-close-packed colloidal crystals, many papers on the fabrication of very large domains of colloidal crystals have been already reported.[8-11] Although (1) the grain size of their crystals are very large, (2) the quality of their crystals is generally fine, and (3) the times for fabrication processes are very short (less than 1 second in the shortest case[8,10]), fabrication of inverse opals from their crystals is difficult.

Although enlargement of widths of columnar grains is also well achieved with the control of nucleation of grains by centrifugation[2,12], further enlargement requires additional developments of growth methods. Several grains were nucleated all at once on the bottom of a flat-bottomed container [Fig. 1(a)]. Growth competitions between adjacent grains occur immediately after the nucleation, and then the average width of the grains soon approaches an asymptotic value. Thus further widening of the grains is difficult in the flat-bottomed container. Initial elimination and subsequent widening of the grains are indispensable for further enlargement of grains.

An inverted-triangle container [Fig. 1(b)] must be useful for the enlargement, since the nucleation of grains is eliminated at an edged bottom, and the grains are widen due to a broadened shape of the container.

In this study, we have tried to enlarge colloidal crystals by centrifugation in an inverted-triangle container, and then compared sizes of the grains in the inverted-triangle container with those in the flat-bottomed container.

An inverted-triangle container (an angle of the edged bottom $\alpha = 90°$) and a flat-bottomed container are schematically shown in Fig. 1. Black polyvinyl-chloride plates were used as spacers for transmitted-light polarization microscopic observation. Aqueous dispersion of silica particles (particle diameter d = 110 nm, volume fraction of the dispersion $\varphi = 0.097$ (18 wt%), Nippon Shokubai) was used without further purification. Crystallization of particles in the dispersion had not occurred for six months during gravitational sedimentation at 1 G (= 9.8 m s$^{-2}$). After gentle agitation by hand, the dispersion (39 and 37 μl) was injected into the containers [(a) a flat-bottomed container and (b) an inverted-triangle container] and the containers were sealed with glass slides and silicone adhesive. Almost close-packed colloidal crystals were fabricated in the containers by centrifugation at 22 m s$^{-2}$ (rotation speed: 190 rpm, radius of rotation: 5.5 cm at the bottoms of the containers) for 30 days at 30°C.

We observed retardation colors of grains in the crystal using a transmitted-light polarization microscope (Olympus, VANOX). Grain boundaries were easily visualized with the retardation colors, since the difference in the retardation colors indicates that in the orientations of grains.[13,14] The containers were sealed in plastic bags with ultra-pure water to avoid evaporation of the dispersion and infiltration of the bubbles into the containers during centrifugation. If a bubble moves up through a grain by centrifugal force, the grain is easily disrupted.

Merged transmission cross-polarized images of colloidal crystals are shown in Fig. 2. Grains of the crystals are distinguished by different retardation colors of grains; grain boundaries are easily confirmed [Fig. 2(e)]. Even though difference in retardation colors between two grains whose orientations similar to each other is very small, a grain boundary shows a darker color than that of the grains, and the color is distinguishable from the colors of the above two grains. Grain boundaries are shown with black dashed lines in Figs. 2(b) and 2(d). Although the structure and orientation of grains, which we have a plan to determine using ultra violet spectra in near future, have not been characterized yet, at least, grain sizes were successfully characterized using the grain boundaries.

The volume of the largest grain in an inverted-triangle container [2.65 × 4 × 0.2 mm$^3$, volume: 1.26 mm$^3$, Fig. 2(c)] was much larger than that in a flat-bottomed container [0.66 × 3.2 × 0.2 mm$^3$, volume: 0.36 mm$^3$, Fig. 2(a)]. Why did the inverted-triangle container enlarge a grain so effectively?

Maximum widths of grains $W_{max}$ with the heights of grains from the bottoms of containers $h$ in a flat-bottomed container and two inverted-triangle containers are shown in Fig. 3. $W_{max}$ in a flat-bottomed container were larger than those in inverted-triangle containers at the bottoms of containers. This is mainly due to faster condensation of particles at the edged bottom in an inverted-triangle container, since the faster condensation results in higher nucleation rates of grains. $W_{max}$ in the inverted-triangle containers become larger than those in the flat-bottomed container when $h$ becomes larger than 1.5 ~ 2.0 mm (Fig. 3).

To clarify the advantages of using an inverted-triangle container, relationships between nucleation rates, particle condensation rates and container shapes should be considered in detail as follows.

(1) Firstly, grains nucleated at the bottom of a flat-bottomed container [Fig. 4(a)-1] and the edged bottom of an inverted-triangle container [Fig. 4(b)-1]. At this stage, particle condensation rates at both bottoms were larger than the critical value for nucleation.

(2) Secondly, in a flat-bottomed container, nucleation was stopped immediately, since particle condensation rates in front of the growth interface of crystals became lower than the critical value for nucleation soon after the interface advanced upward and approached the center of rotation. Thus, only the growth of the columnar crystals followed; no additional nucleation occurred [Fig. 4(a)-2]. Whereas, in an inverted-triangle container, nucleation occurred for a while after the successive growth of the crystals started onto the nuclei at the bottom [Fig. 4(b)-2]. Although centrifugal acceleration at the bottom of a flat-bottomed container (22 m s$^{-2}$) was equal to that of an inverted-triangle container (22 m s$^{-2}$), i.e., particle condensation rates in front of the growth interfaces of colloidal crystals at the same height from the bottoms of containers were the same for both containers, the condensation rates around an edge of the growth interface in an inverted-triangle container were still larger than the critical value for nucleation. This is due to the additional condensation of particles along the slope of a side wall of an inverted-triangle container as schematically shown in Fig. 4(c).

(3) Finally, nucleation at both the edges of the growth interface of crystals in an inverted-triangle container was stopped when the particle condensation rates at the edge became lower than the critical value for nucleation [Fig. 4(b)-3]. From this point forward outmost grains in the inverted-triangle container could be widened outward freely, whereas the grains in a flat-bottomed container could not.

Free space for the outmost grains described in the above (3) is the most important advantage of using an inverted-triangle container, whereas the additional nucleation owing to the additional condensation of particles along the slope of a side wall [in the above (2)] is an undesirable disadvantage. A balance between the advantage and disadvantage is easily controlled with centrifugal acceleration. In this study, we could successfully control the balance, and effectively enlarge a grain in an inverted-triangle container as a result.

We stress here that a centrifugation method was adopted not for the acceleration of crystallization processes but for the precise control of nucleation rates. In this study, rotation speed was set at 190 rpm (22 m s$^{-2}$ at the bottom of containers), and it took 30 days to obtain the crystals as shown in Fig. 2, since slower centrifugation results in the reduction of nucleation rates.[12] Although Nishijima et al. fabricated 100-μm-thick, uniform, and close-packed colloidal crystals by centrifugation at 41000 m s$^{-2}$ (4200 G) in 5 min[15], they did not say that the obtained films were single crystals, or not.

We have to note that the flatness of the container wall was not high as shown in Fig. 2, since the containers were made by hand. Fortunately, rough surface of the container wall have not induced undesirable nucleation during the growth of the largest grain in Fig. 2(c).

Although we successfully achieved the enlargement of grains by centrifugation using an inverted-triangle container, the quality of grains has not been controlled yet. The second largest grain in the inverted-triangle container in Fig. 2(c) contains striations inside. The striations indicate stacking disorders (intrinsic or extrinsic stacking faults, or twin band structures[13,14]). To reduce such striations we are planning to apply stronger centrifugal forces to the obtained crystals. Centrifugation would be the best choice for annealing the crystals, since we have already fabricated large grains by centrifugation, and gravitational forces is known to reduce stacking disorders in hard-sphere crystals by Monte Carlo simulations.[16,17] Actually, we have already succeeded in reducing the number of striations by applying larger centrifugal acceleration to the obtained crystals as a preliminary result.[18] Precise control of a "gravitational annealing" will be the future work.

Crystal structure analysis by UV microscopic spectroscopy is remained for the future work. The precise analysis of the structure and orientation of grains as reported in ref. 19 and 20 will help us to consider the deeper understandings of mechanisms of enlargement of grains.

We successfully obtained large grains of silica colloidal crystals with an inverted-triangle container by centrifugation. The largest volume of grains in an inverted-triangle container was much larger than that in a flat-bottomed container. The number of grains was eliminated at the edged bottom of an inverted-triangle container. Outmost grains of the inverted-triangle container were freely and effectively widened horizontally owing to the inverted-triangle shape of the container accompanied by the upward advancement of the growth interface of crystals.

Fig. 1. Schematic illustrations of (a) a flat-bottomed container and (b) an inverted-triangle internal-shaped container with an edged bottom ($\alpha = 90°$). Black polypropylene plates (thickness: 0.2 mm) are used as spacers between glass slides.

Fig. 2. (Color online) Merged cross-polarized images of the colloidal crystals in (a) a flat-bottomed container and (c) an inverted-triangle internal-shaped container and images with grain boundaries (black dashed lines) in (b) a flat-bottomed container and (d) an inverted-triangle internal-shaped container. (e) Magnified image of grain boundaries shown in (c), and (f) shows the image with grain boundaries (black dashed lines).

Fig. 3. Widths of the largest grains in two inverted-triangle containers and a flat-bottomed container with the growth of crystals.

Fig. 4. Schematic illustrations of nucleation and growth of grains of colloidal crystals in a flat-bottomed container (a) and an inverted-triangle container (b). A schematic illustration of additional condensation of particles around an edge of the growth interface of colloidal crystals in an inverted-triangle container (c).

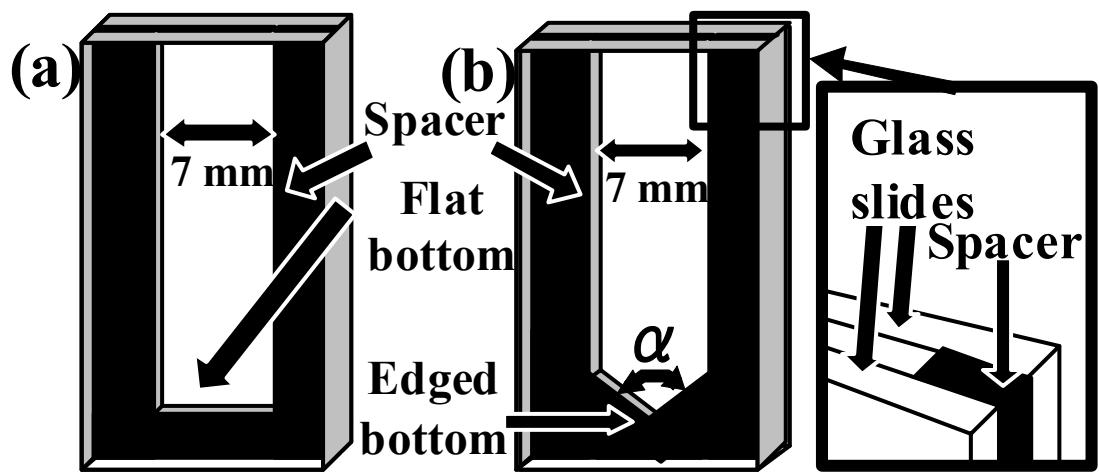

Fig. 1

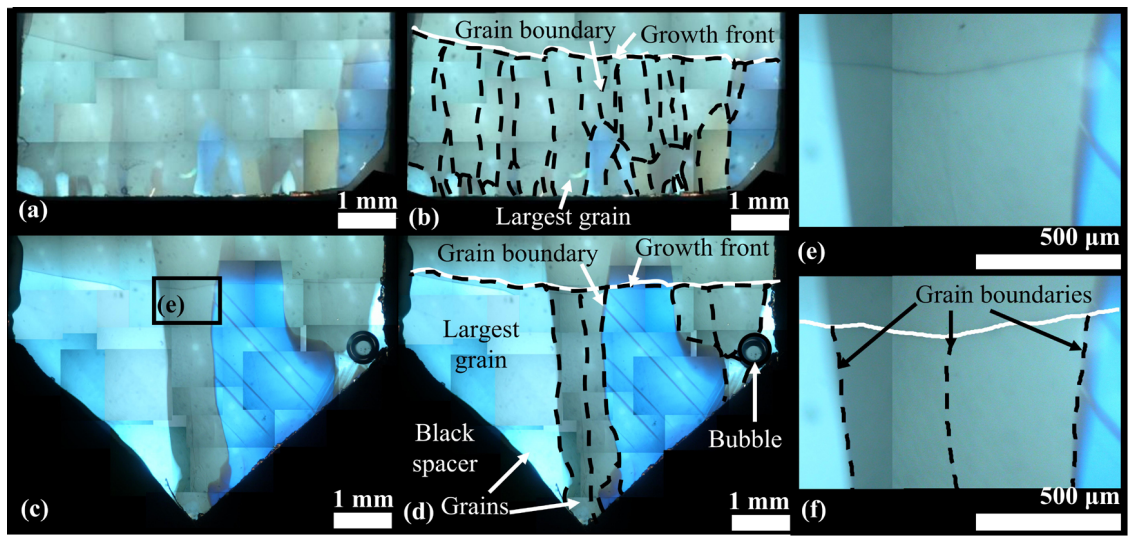

Fig. 2

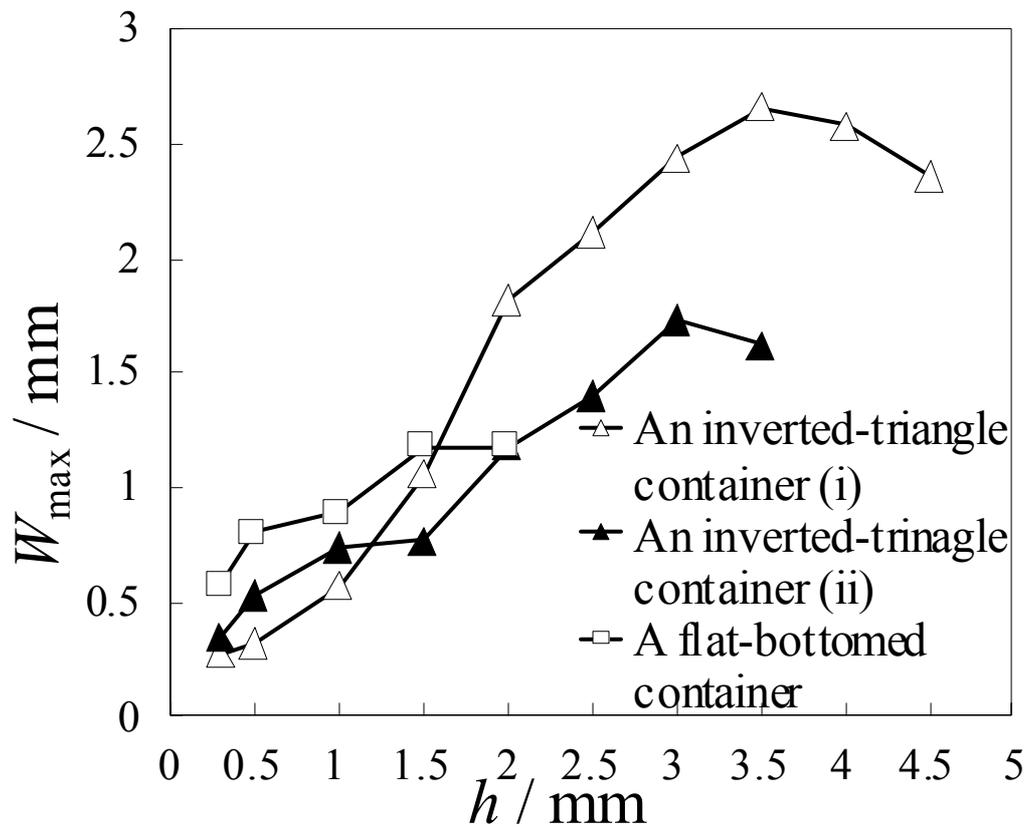

Fig. 3

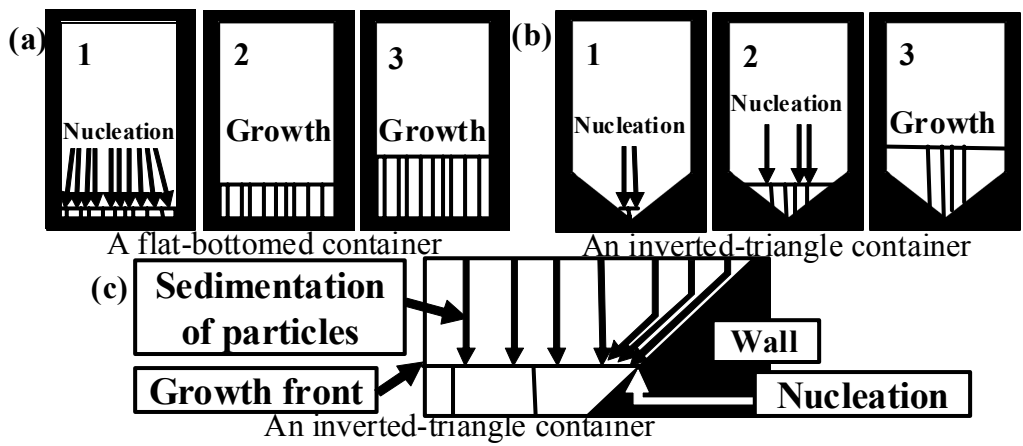

Fig. 4